\documentclass[12pt]{iopart}

\usepackage{iopams}
\usepackage{graphicx}
\usepackage{longtable}
\begin{document}

\title[Single-$\Lambda$ Hypernuclei
in the Relativistic Mean-Field Theory]{Single-$\Lambda$ Hypernuclei
in the Relativistic Mean-Field Theory with parameter set FSU}

\author{Renli Xu$^{1}$, Chen Wu$^{2}$, Zhongzhou Ren$^{1,3,4}$}
\address{$^{1}$Department of Physics, Nanjing University,
 Nanjing 210093, China}
\address{$^{2}$Shanghai Institute of Applied Physics, Chinese Academy of Sciences, Shanghai 201800, China}
\address{$^{3}$Center of Theoretical Nuclear Physics, National
 Laboratory of Heavy-Ion Accelerator, Lanzhou 730000, China}
\address{$^{4}$Kavli Institute for Theoretical Physics China, Beijing 100190, China}
\ead{wuchenoffd@gmail.com}
\begin{abstract}

In this work, we conduct a study on the properties of
single-$\Lambda$ hypernuclei in the relativistic mean-field theory with the parameter set FSU,
where the isoscalar-isovector coupling has been included to soften the symmetry energy.
In our model, a tensor coupling between $\omega$ and $\Lambda$ is employed,
which is found to be essential
in obtaining the small spin-orbit interaction in single-$\Lambda$ hypernuclei. 
Our calculated values of $\Lambda$ single-particle
energies are in good agreement with the known experimental data.
As a comparison to other existing parameter sets,
calculations are also carried out by using parameterizaions such as NL3 and NL3$^*$.
The results show that the parameter set FSU is as
successful as those of NL3 and NL3$^*$ in terms of reproducing the properties of
single-$\Lambda$ hypernuclei.
\end{abstract}

\submitto{\jpg}
\maketitle

\section{Introduction}

The Relativistic Mean-Field Theory (RMF) has been widely used to
study the properties of finite nuclei and nuclear matter and
it has 
achieved great success \cite{Ring,Lalazissis,Sugahara1,Ren}.
With several parameters, the RMF theory gives good description of the 
properties of nuclei, such as binding energy, rms radii, and single-particle energy levels.
It can also be used to
investigate the properties of $\Lambda$ hypernuclei as well as multi-lambda hypernuclei\cite{Rufa}.
Since there exist only a few experimental data for hyperon-nucleon (YN) and
hyperon-hyperon (YY) interactions, the study of hypernuclei 
provides important information 
on the YN and YY interactions.
In fact, exploring the nuclear systems with strangeness has long attracted lots of
attentions 
and theoretical progresses have been made over the past years\cite{Rufa,Mares1,Win,Vretenar,Tsushima1,Shen1}.
On the experimental side, the recent observation of an antimatter hypernucleus
$\tiny{^{3}_{\bar{\Lambda}}}$$\mathrm{\bar{H}}$ in relativistic
heavy-ion collisions 
by STAR Collaboration\cite{Star} has aroused much interest.
Besides its astrophysical and cosmological implications,
strange nuclear physics is indeed an attractive topic by itself.

We know that baryons interact via the exchange of mesons in the RMF framework.
In $\Lambda$ hypernuclei, the baryons include nucleons and $\Lambda$ hyperons.
For the nucleonic sector, we will make use of the
parameter set FSU, which was proposed
by Todd-Rutel and Piekarewicz in 2005.
The parameter set NL3\cite{Lalazissis1, Lalazissis2} and recently the improved
version NL3$^*$\cite{Lalazissis3} are considered to be very successful for finite nuclei.
For comparison purpose, calculations will also be carried out using these parameterizations.
The FSU parameter set predicts a compression modulus for symmetric nuclear matter
of $K = 230$ MeV and a neutron skin in $\tiny{^{208}}$Pb with a thickness
of $R_n - R_p = 0.21$ fm \cite{Todd-Rutel}, while the thickness of neutron skin
in $\tiny{^{208}}$Pb was estimated 
to be 0.28 fm by the NL3\cite{Lalazissis1,Horowitz}.
The additional isoscalar-isovector
coupling ($\Lambda_\mathrm{v}$) term in FSU is introduced to soften
the symmetry energy at high densities \cite{Todd-Rutel}.
The validity of parameter set FSU has been examined by calculating the properties of nuclear
matter, finite nuclei in some previous works\cite{Sheng,Wu}, and our aim in
this work is to use the parameter set FSU as well as the NL3 and
the NL3$^*$ to study the properties of single-$\Lambda$ hypernuclei.

It is widely believed that the spin-orbit splitting in
$\Lambda$ hypernuclei is quite small. The tensor coupling between
$\omega$ and $\Lambda$ acts as a spin-orbit interaction,
and has an important impact
on the spin-orbit splitting of hyperons, 
as observed by Jennings\cite{Jennings}.
Following this line of thought, in our calculation we evaluate 
the effective potenials of hyperon in hypernuclei, particularly the
spin-orbit potentials, in order to study the effect of
the tensor coupling on the spin-orbit splitting of $\Lambda$ hyperon in hypernuclei.

The organization of this paper is as follows.
In Sec. 2, we outline the theoretical framework of $\Lambda$ hypernuclei in RMF theory,
where tensor coupling is introduced in addition to the FSU parameter set.
The model parameters and numerical results are presented and discussed in Sec. 3.
A summary is given in Sec. 4.

\section{Formulas of the RMF model with hyperon}

In the RMF theory, the nuclear interaction is usually described by
the exchange of three mesons: the isoscalar meson $\sigma$, the
isoscalar-vector meson $\omega$, and the isovector-vector meson
$\rho$. There exists electromagnetic interaction between protons in
nuclei, so the photon $A$ is included.
The baryons involved in the
present work are nucleons and $\Lambda$ hyperon. The $\Lambda$
hyperon is a charge neutral and isoscalar particle so that it does
not couple to isovector-vector $\rho$ and photon A. The effective
Lagrangian density is in the following form:
\begin{eqnarray}
\mathcal{L} &=&
\bar{\psi}[\gamma^{\mu}(i\partial_\mu-g_\omega\omega_\mu-\frac{g_\rho}{2}\tau\cdot\rho_\mu-
\frac{e}{2}(1+\tau_3)A_\mu)-M_N^*]\psi
\nonumber\\&&+\bar{\psi}_\Lambda[\gamma^{\mu}(i\partial_\mu-g^\Lambda_\omega\omega_\mu
)-M_{\Lambda}^*+\frac{f_\omega^\Lambda}{2M_\Lambda}\sigma^{\mu\nu}\partial_\nu\omega_\mu]\psi_\Lambda
\nonumber\\&&+\frac{1}{2}\partial^\mu\sigma\partial_\mu\sigma-\frac{1}{2}m^2_\sigma\sigma^2
-\frac{1}{4}V^{\mu\nu}V_{\mu\nu}+\frac{1}{2}m^2_\omega\omega^\mu\omega_\mu
\nonumber\\&&-\frac{1}{4}b^{\mu\nu}b_{\mu\nu}+\frac{1}{2}m^2_\rho\rho^\mu\rho_\mu-
\frac{1}{4}F^{\mu\nu}F_{\mu\nu}-U_{eff}(\sigma,\omega^\mu,\rho^\mu)
\end{eqnarray}
where the effective mass for the $\Lambda$ hyperon and nucleon are
defined as: $M_N^*=M_N-g_\sigma\sigma$,
$M_\Lambda^*=M_\Lambda-g_\sigma^\Lambda\sigma$. The strength tensors
of the vector mesons and electromagnetic field are defined as:
$V_{\mu\nu}=\partial_\mu\omega_\nu-\partial_\nu\omega_\mu$,
$b_{\mu\nu}=\partial_\mu\rho_\nu-\partial_\nu\rho_\mu$,
$F_{\mu\nu}=\partial_\mu A_\nu-\partial_\nu A_\mu$. The
self-interacting terms of $\sigma$, $\omega$ mesons and the
isoscalar-isovector intertaction one are taken as:
\begin{eqnarray}
U_{eff}(\sigma,\omega^\mu,\rho^\mu)&=&
\frac{\kappa}{3!}(g_\sigma\sigma)^3+\frac{\lambda}{4!}(g_\sigma\sigma)^4-\frac{\zeta}{4!}(g^2_\omega\omega_\mu\omega^\mu)^2\nonumber\\
&&-\Lambda_\mathrm{v}(g^2_\rho\rho_\mu\rho^\mu)(g^2_\omega\omega_\mu\omega^\mu).
\end{eqnarray}
Here, the isoscalar meson self-interactions (via $\kappa$,
$\lambda$, and $\zeta$ ) are necessary for the appropriate equation
of state of symmetric nuclear matter \cite{Lalazissis1,Toki}. The
new additional isoscalar-isovector coupling ($\Lambda_\mathrm{v}$)
term is used to modify the density dependence of the symmetry energy
and the neutron skin thicknesses of heavy nuclei
\cite{Todd-Rutel,Sheng}. The $\psi$ and $\psi_\Lambda$ are the Dirac
spinors for nucleons and $\Lambda$ hyperon.

Using the mean-field approximation, the Dirac equations for nucleons
and $\Lambda$ hyperon have the following form:
\begin{eqnarray}
\lefteqn{[i\gamma^{\mu}\partial_\mu-(M_N-g_\sigma\sigma)-g_\omega\gamma^0\omega_0-\frac{g_\rho}{2}\gamma^0\tau_3\rho_0-\frac{e}{2}\gamma^0(1+\tau_3)A_0]\psi=0,}\nonumber\\
\lefteqn{[i\gamma^{\mu}\partial_\mu-(M_\Lambda-g_\sigma^\Lambda\sigma)-g_\omega^\Lambda\gamma^0\omega_0+\frac{f_\omega^\Lambda}{2M_\Lambda}\sigma^{0i}\partial_i\omega]\psi_\Lambda=0}
\end{eqnarray}

The Klein-Gordon equations for the mesons and photon can be written
as\begin{eqnarray}
\lefteqn{\lefteqn(-\Delta+m^2_\sigma)\sigma(\textbf{r})=g_\sigma\rho_s(\textbf{r})+g^\Lambda_\sigma\rho^\Lambda_s(\textbf{r})
-\frac{\kappa}{2}g^3_\sigma\sigma^2(\textbf{r})-\frac{\lambda}{6}g^4_\sigma\sigma^3(\textbf{r}),}\nonumber\\
\lefteqn{\lefteqn(-\Delta+m^2_\omega)\omega_0(\textbf{r})=g_\omega\rho_v(\textbf{r})+g^\Lambda_\omega\rho^\Lambda_v(\textbf{r})-\frac{\zeta}{6}g^4_\omega\omega^3_0(\textbf{r})}\nonumber\\
\lefteqn{\kern3.2cm-2\Lambda_\mathrm{v}g^2_\rho
g^2_\omega\rho^2_0(\textbf{r})\omega_0(\textbf{r})
+\frac{f_\omega^\Lambda}{2M_\Lambda}\rho_0^T(\textbf{r}),}
\lefteqn\nonumber\\
\lefteqn{(-\Delta+m^2_\rho)\rho_0(\textbf{r})=\frac{g_\rho}{2}\rho_3(\textbf{r})-2\Lambda_\mathrm{v}
g^2_\rho g^2_\omega\omega^2_0(\textbf{r})\rho_0(\textbf{r}),}
\lefteqn\nonumber\\
\lefteqn{-\Delta A_0(\textbf{r})=e\rho_p(\textbf{r}).}
\end{eqnarray}
where
$\rho_s(\textbf{r})(\rho^\Lambda_s(\textbf{r})),\rho_v(\textbf{r})(\rho^\Lambda_v(\textbf{r})),
\rho_0^T(\textbf{r}),\rho_3(\textbf{r})$ and $\rho_p(\textbf{r})$
are the scalar, vector, tensor, third component of isovector, and
proton densities, respectively. Eqs.(3) and (4) can be
self-consistently solved by iteration. We focus our study on the
spherical case, and use the BCS theory to calculate the pairing
contribution for open shell nuclei. The detail of the solution can
be easily found in the literatures \cite{Ring,Gambhir}, and it is
not reiterated here.

\begin{center}
\begin{indented}
\lineup\item[]
\begin{longtable}{@{}lcccccccccc}
\caption{\label{table1}The parameter sets NL3, NL3$^*$, and FSU. The nucleon and rho
masses are respectively set as $M_N$=939 MeV, $m_\rho$=763 MeV.}\\
\br
&$m_\sigma$(MeV)&$m_\omega$(MeV)&$g^2_\sigma$&$g^2_\omega$&$g^2_\rho$&$\kappa$(MeV)&$\lambda$&$\zeta$&$\Lambda_\mathrm{v}$\\
\mr
\endfirsthead

\multicolumn{10}{c}%
{{\bfseries \tablename\ \thetable{} -- continued from previous page}} \\
\br
&$m_\sigma$(MeV)&$m_\omega$(MeV)&$g^2_\sigma$&$g^2_\omega$&$g^2_\rho$&$\kappa$(MeV)&$\lambda$&$\zeta$&$\Lambda_\mathrm{v}$\\
\mr
\endhead
\br \multicolumn{10}{r}{{(Continued on next page)}} \\
\endfoot
\br
\endlastfoot
NL3&508.1940&782.501&104.3871&165.5854&80.0667&3.8599&--0.0159&0.000&0.000\\
NL3$^*$&502.5742&782.600&101.8969&164.0064&83.7152&4.1474&--0.0174&0.000&0.000\\
FSU&491.5000&782.500&112.1996&204.5469&138.4701&1.4203&0.0238&0.060&0.030\\
\end{longtable}
\end{indented}
\end{center}

\section{Numerical results and discussion}
In our calculation, we use the parameter sets NL3, NL3$^*$ and FSU
\cite{Todd-Rutel,Fattoyev} for the parameters of the nucleonic
sector, whose are tabulated in \tref{table1}. At first, we define
the ratios of the meson-hyperon couplings to the meson-nucleon
couplings, $R_\sigma=g_\sigma^\Lambda/g_\sigma$ and
$R_\omega=g_\omega^\Lambda/g_\omega$. We take the naive quark model
value for the relative $\omega$ coupling as $R_\omega = 2/3$
\cite{Shen,Keil}. 
To reproduce the experimental $\Lambda$
binding energies of single-$\Lambda$ hypernuclei, we adopt the
relative $\sigma$ coupling as $R_\sigma=0.619$ for the FSU,
$R_\sigma=0.620$ for the NL3$^*$ and $R_\sigma=0.621$ for the NL3 parameter set.
We take the experimental nucleon mass as $M_N=939$ MeV and lambda
hyperon mass as $M_\Lambda=1116$ MeV, respectively.
\begin{center}
\begin{footnotesize}
\lineup\item[]
\begin{longtable}{*{6}{c @{\extracolsep\fill}}}
\caption{\label{table2}$\Lambda$ single-particle energies(in MeV)
for nuclear core plus one $\Lambda$ configuration. The values for
$\tiny{^{16}_{~\Lambda}}$O, $\tiny{^{40}_{~\Lambda}}$Ca,
$\tiny{^{48}_{~\Lambda}}$Ca, $\tiny{^{51}_{~\Lambda}}$V,
$\tiny{^{89}_{~\Lambda}}$Y, $\tiny{^{139}_{~~\Lambda}}$La and
$\tiny{^{208}_{~~\Lambda}}$Pb are calculated. The available
experimental data denoted by a are taken from Ref.\cite{Usmani} and
denoted by b are taken from
Ref.\cite{Hashimoto1} .}\\
\br
&$\Lambda$ state&NL3&NL3$^*$&FSU&Expt.\\
\mr
\endfirsthead

\multicolumn{6}{c}%
{{\bfseries \tablename\ \thetable{} -- continued from previous page}} \\
\br
&$\Lambda$ state&NL3&NL3$^*$&FSU&Expt.\\
\mr
\endhead
\br \multicolumn{6}{r}{{(Continued on next page)}}\\
\endfoot
\br
\endlastfoot
$\tiny{^{16}_{~\Lambda}}$O&$1s_{1/2}$&--12.91&--12.64&--12.88&--12.5$\pm$0.35$^a$\\
&$1p_{3/2}$&--2.96&--2.88&--2.49\\
&$1p_{1/2}$&--2.73&--2.66&--2.26&--2.5$\pm$0.5$^a$\\
\hline
$\tiny{^{40}_{~\Lambda}}$Ca&$1s_{1/2}$&--19.44&--19.28&--19.33&--18.7$\pm$1.1$^a$\\
&$1p_{3/2}$&--10.59&--10.39&--10.26&\\
&$1p_{1/2}$&--10.34&--10.14&--10.04&--11.0$\pm$0.6$^a$\\
&$1d_{5/2}$&--2.62&--2.52&--2.07&\\
&$2s_{1/2}$&--2.40&--2.41&--2.10&\\
&$1d_{3/2}$&--2.35&--2.26&--1.81&--1.0$\pm$0.5$^a$\\
\hline
$\tiny{^{48}_{~\Lambda}}$Ca&$1s_{1/2}$&--21.08&--20.88&--20.87&\\
                           &$1p_{3/2}$&--12.53&--12.31&--12.22&\\
                           &$1p_{1/2}$&--12.31&--12.09&--12.03&\\
                           &$1d_{5/2}$&--4.37&--4.23&--3.85&\\
                           &$2s_{1/2}$&--3.64&--3.63&--3.21&\\
                           &$1d_{3/2}$&--4.08&--3.95&--3.57&\\
\hline
$\tiny{^{51}_{~\Lambda}}$V&$1s_{1/2}$&--21.82&--21.64&--21.39&--19.9$\pm$1.0$^a$\\
                          &$1p_{3/2}$&--13.29&--13.06&--12.87&\\
                          &$1p_{1/2}$&--13.08&--12.85&--12.69&\\
                          &$1d_{5/2}$&--4.98&--4.83&--4.43&\\
                          &$2s_{1/2}$&--4.06&--4.05&--3.56&\\
                          &$1d_{3/2}$&--4.67&--4.53&--4.14&--4.0$\pm$0.5$^a$\\
\hline
$\tiny{^{89}_{~\Lambda}}$Y&$1s_{1/2}$&--24.05&--23.90&--23.66&--23.1$\pm$0.5$^b$\\
                          &$1p_{3/2}$&--17.47&--17.25&--17.13&\\
                          &$1p_{1/2}$&--17.34&--17.11&--17.03&--16.5$\pm$4.1$^b$\\
                          &$1d_{5/2}$&--10.47&--10.26&--10.02&\\
                          &$2s_{1/2}$&--8.82&--8.73&--8.21&\\
                          &$1d_{3/2}$&--10.23&--10.02&--9.80&--9.1$\pm$1.3$^b$\\
                          &$1f_{5/2}$&--3.22&--3.08&--2.66&--2.3$\pm$1.2$^b$\\
\hline
$\tiny{^{139}_{~\Lambda}}$La&$1s_{1/2}$&--25.29&--25.09&--24.94&--24.5$\pm$1.2$^b$\\
                            &$1p_{3/2}$&--20.37&--20.14&--20.06&\\
                            &$1p_{1/2}$&--20.32&--20.09&--20.01&--20.4$\pm$0.6$^b$\\
                            &$1d_{5/2}$&--14.65&--14.41&--14.29&\\
                            &$2s_{1/2}$&--12.51&--12.34&--12.04&\\
                            &$1d_{3/2}$&--14.51&--14.27&--14.17&--14.3$\pm$0.6$^b$\\
                            &$1f_{7/2}$&--8.50&--8.29&--8.04\\
                            &$1f_{5/2}$&--8.26&--8.05&--7.82&--8.0$\pm$0.6$^b$\\
                            &$1g_{7/2}$&--1.98&--1.84&--1.44&--1.5$\pm$0.6$^b$\\
\hline
$\tiny{^{208}_{~\Lambda}}$Pb&$1s_{1/2}$&--26.69&--26.52&--26.26&--26.3$\pm$0.8$^b$\\
                            &$1p_{3/2}$&--22.67&--22.46&--22.28&\\
                            &$1p_{1/2}$&--22.63&--22.41&--22.24&--21.9$\pm$0.6$^b$\\
                            &$1d_{5/2}$&--17.89&--17.65&--17.51&\\
                            &$2s_{1/2}$&--15.91&--15.73&--15.51&\\
                            &$1d_{3/2}$&--17.79&--17.55&--17.42&--16.8$\pm$0.7$^b$\\
                            &$1f_{7/2}$&--12.60&--12.36&--12.17&\\
                            &$1f_{5/2}$&--12.42&--12.18&--12.01&--11.7$\pm$0.6$^b$\\
                            &$1g_{7/2}$&--6.78&--6.58&--6.28&--6.6$\pm$0.6$^b$\\

\end{longtable}
\end{footnotesize}
\end{center}
In the present calculation, we adopt the quark model value of the tensor coupling,
$f_\omega^\Lambda=-g_\omega^\Lambda$ \cite{Jennings,Sugahara}. The
center-of-mass correction to the binding energy is given as
$E_{c.m.}=-\frac{3}{4}\times41A^{-1/3}$MeV.

The $\Lambda$ binding energy is a very important quantity for the
investigation of the properties of $\Lambda$ hypernuclei. We first
calculate the $\Lambda$ single-particle energies for some typical
spherical hypernuclei. The results are shown in table 2. By putting
the $\Lambda$ particle either in the 1$s_{1/2}$ orbit or in one of
the other orbits, we can get the corresponding $\Lambda$
single-particle energy. One can easily see from table 2 that the
$\Lambda$ binding energies in our calculation are in good agreement
with the experimental value and the parameter set FSU is as good as
those of NL3 and NL3$^*$ for the binding
energies of the $\Lambda$ hypernuclei. We also find that when the
nuclear number increases, the $\Lambda$ binding energy in $\Lambda$
hypernuclei for $1s_{1/2}$ state becomes smaller, and finally approaches
the binding energy for $\Lambda$ bounded in nuclear matter with
$\epsilon_\Lambda=-28$ MeV \cite{Wu,Millener}.

The small spin-orbit splittings of $\Lambda$ hyperon are also
obtained in our calculation. 
As usually undertood, the spin-orbit
splitting of the p orbit is quite small, --0.8 MeV $\leq
\delta_\Lambda$ $\leq$ 0.2 MeV \cite{Hashimoto1}, where
$\delta_\Lambda=\epsilon_\Lambda(p_{1/2})-\epsilon_\Lambda(p_{3/2})$.
In Ref. \cite{Hashimoto1}, the energy difference between the $2_1^+$
and $0_1^+$ states of $\tiny{^{16}_{~\Lambda}}$O is obtained to be
0.04 $\pm$ 0.32 MeV. From \tref{table2}, we can see that the
$\delta_\Lambda$ of 1p orbit for $\tiny{^{16}_{~\Lambda}}$O is about
--0.23 MeV for NL3, NL3$^*$ and FSU,  which is consistent with the
results in Ref. \cite{Hashimoto1}. 
As a comparison, we also calculate
the $\Lambda$ binding energies without tensor coupling between
$\omega$ and $\Lambda$, where the result of $\delta_\Lambda$ of 1p
orbit for $\tiny{^{16}_{~\Lambda}}$O is --1.41 MeV with the FSU,
which is much bigger than case with tensor coupling.
Therefore we can conclude that the tensor coupling plays an important role in the
spin-orbit interaction in single-$\Lambda$ hypernuclei.

Another interesting result concerns the $\Lambda$ binding energies of
$2s_{1/2}$ and $1d_{3/2}$. The $\Lambda$ binding energy of
$2s_{1/2}$ is bigger than $1d_{3/2}$ for
$\tiny{^{40}_{~\Lambda}}$Ca, while it is smaller in the case of
$\tiny{^{48}_{~\Lambda}}$Ca, $\tiny{^{51}_{~\Lambda}}$V, $\tiny{^{89}_{~\Lambda}}$Y,
$\tiny{^{139}_{~~\Lambda}}$La and $\tiny{^{208}_{~~\Lambda}}$Pb.
What makes it important
is that the results calculated in all of the three parameter sets ( NL3, NL3$^*$, FSU)
show the same phenomena.
This can be understood as a result of the presence of tensor coupling.
As one can see from table 2 that 
the spin-orbit splitting of 1d level
of $\Lambda$ hyperon reduces as the nuclear number increases.
As a consequence, the energy level of the $\Lambda$ orbit $1d_{3/2}$ is
pulled down, and the difference of energy levels between $2s_{1/2}$ and
$1d_{3/2}$ switches its sign.

\begin{figure}[htb]
\centering
\includegraphics[width=15cm]{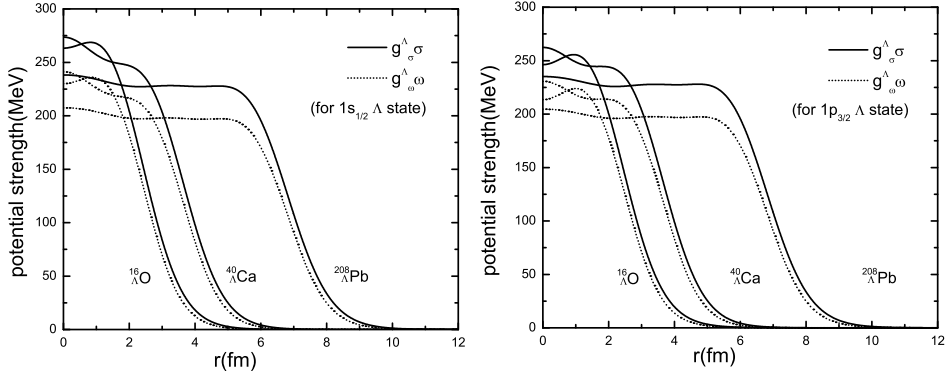}
\caption{Scalar and vector potentials of $\Lambda$ in
$\tiny{^{16}_{~\Lambda}}$O and $\tiny{^{40}_{~\Lambda}}$Ca and
$\tiny{^{208}_{~~\Lambda}}$Pb, respectively. The results of
$1p_{3/2}$ $\Lambda$ state are also shown for
comparison. The using parameter set is FSU.}\label{fig1}
\end{figure}

We know that the spin-orbit interaction of baryons contain
contributions from the tensor potential and the derivative of the
sum of the scalar and vector potentials. By recasting the Dirac
equation in Schr\"{o}dinger equivalent form, we can obtain the
spin-orbit potential of $\Lambda$ hyperon $V_{ls}^\Lambda l\cdot s$,
with \cite{Mares}.
\begin{eqnarray}
\lefteqn{V_{ls}^\Lambda l\cdot
s=\frac{1}{2M_{eff}^2}[\frac{1}{r}(g_\omega^\Lambda\frac{\partial \omega_0}{\partial r}+g_\sigma^\Lambda\frac{\partial \sigma}{\partial r}+2f_{\omega\Lambda}
\frac{M_{eff}}{M_\Lambda}\frac{\partial \omega_0}{\partial r})]l\cdot s,}\\
\lefteqn{M_{eff}=M_\Lambda-\frac{1}{2}(g_\omega^\Lambda\omega_0+g_\sigma^\Lambda\sigma).}
\end{eqnarray}

\begin{figure}[tbp]
\begin{center}
\includegraphics[width=15cm]{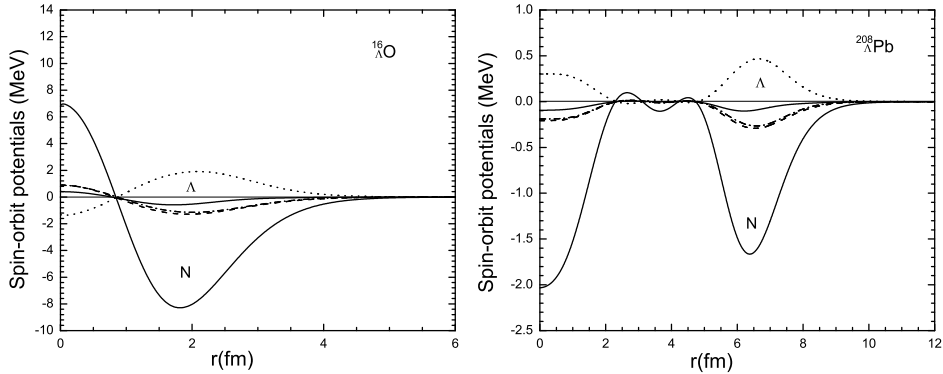}\end{center}
\caption{Spin-orbit potentials of baryons in
$\tiny{^{16}_{~\Lambda}}$O and $\tiny{^{208}_{~~\Lambda}}$Pb. The
solid curves are the total spin-orbit potentials for a proton and a
$\Lambda$. The dotted, dash-dotted and dashed ones are those for a
$\Lambda$ produced by the tensor, vector and scalar potentials,
respectively. All cases are for the $1s_{1/2}$ $\Lambda$ state. The using parameter set is FSU.}
\label{fig2}
\end{figure}

The scalar and vector potentials of the $\Lambda$ hyperon for
$\tiny{^{16}_{~\Lambda}}$O, $\tiny{^{40}_{~\Lambda}}$Ca and
$\tiny{^{208}_{~~\Lambda}}$Pb are plotted in Fig.1, where the solid
curves are for scalar potentials and the dotted ones for vector
potentials, the 1$p_{3/2}$ $\Lambda$ state are also shown for
comparison. From Fig.1, we can see that the difference between the
scalar and vector potentials near the center of the hypernuclei is
typically $\sim 30-35$ MeV, which is consistent with the result
calculated by Ma et al. \cite{Ma}. Additionally, the scalar and
vector potentials near the center of the hypernuclei in $1p_{3/2}$
$\Lambda$ state is smaller than those in $1s_{1/2}$ $\Lambda$ state,
which cause the root-mean-square (rms) radius  of the $\Lambda$
($r_{\Lambda}$) in $1p_{3/2}$ state become bigger than in $1s_{1/2}$
state.

The spin-orbit potentials for $\tiny{^{16}_{~\Lambda}}$O and
$\tiny{^{208}_{~~\Lambda}}$Pb are shown in Fig.2, where dash-dotted,
dashed, dotted curves are produced by first, second and third terms
in Eq.(5), respectively. From Fig.2, we can see that the tensor
potentials is comparable in magnitude with scalar and vector
potentials, however, it has a negative sign. As a result, the
spin-orbit term for $\Lambda$ became very small. It is also seen
from Fig.2 that the spin-orbit splitting of proton in
$\tiny{^{16}_{~\Lambda}}$O is much bigger than that in
$\tiny{^{208}_{~~\Lambda}}$Pb. For example,
$\delta_P$$(\delta_P=\epsilon_P(p_{1/2})-\epsilon_P(p_{3/2}))$ of 1p
state for $\tiny{^{16}_{~\Lambda}}$O is --6.99 MeV, but it is
only --0.61 MeV for $\tiny{^{208}_{~~\Lambda}}$Pb in our
calculation.

\begin{center}
\begin{indented}
\lineup\item[]
\begin{longtable}{*{7}{c @{\extracolsep\fill}}}
\caption{\label{table3}Binding energy per baryon, E/A(in MeV), rms
charge radius($r_{c}$), and rms radii of the $\Lambda$($r_\Lambda$),
neutron($r_n$), proton($r_p$)(in fm). The results are calculated in FSU.}\\
\br
&$\Lambda$ state&$E/A$&$r_{c}$&$r_{\Lambda}$&$r_n$&$r_p$\\
\endfirsthead

\multicolumn{7}{c}%
{{\bfseries \tablename\ \thetable{} -- continued from previous page}} \\
\br
&$\Lambda$ state&$E/A$&$r_{c}$&$r_{\Lambda}$&$r_n$&$r_p$\\
\mr
\endhead
\br \multicolumn{7}{r}{{(Continued on next page)}} \\
\endfoot
\br
\endlastfoot
\hline
$\tiny{^{17}_{~\Lambda}}$O&$1s_{1/2}$&--8.25&2.68&2.42&2.53&2.56\\
$\tiny{^{17}_{~\Lambda}}$O&$1p_{3/2}$&--7.65&2.69&3.69&2.54&2.57\\
$\tiny{^{16}}$O&&--7.96&2.69&&2.54&2.56\\
\hline
$\tiny{^{41}_{~\Lambda}}$Ca&$1s_{1/2}$&--8.78&3.43&2.66&3.28&3.33\\
$\tiny{^{41}_{~\Lambda}}$Ca&$1p_{3/2}$&--8.56&3.44&3.49&3.29&3.34\\
$\tiny{^{40}}$Ca&&--8.52&3.43&&3.29&3.34\\
\hline
$\tiny{^{49}_{~\Lambda}}$Ca&$1s_{1/2}$&--8.82&3.46&2.74&3.57&3.37\\
$\tiny{^{49}_{~\Lambda}}$Ca&$1p_{3/2}$&--8.64&3.47&3.49&3.57&3.37\\
$\tiny{^{48}}$Ca&&--8.57&3.47&&3.57&3.37\\
\hline
$\tiny{^{90}_{~\Lambda}}$Y&$1s_{1/2}$&--8.85&4.24&3.16&4.27&4.16\\
$\tiny{^{90}_{~\Lambda}}$Y&$1p_{3/2}$&--8.77&4.24&3.85&4.27&4.16\\
$\tiny{^{89}}$Y&&--8.68&4.24&&4.27&4.16\\
\hline
$\tiny{^{140}_{~\Lambda}}$La&$1s_{1/2}$&--8.53&4.86&3.63&4.96&4.79\\
$\tiny{^{140}_{~\Lambda}}$La&$1p_{3/2}$&--8.49&4.86&4.29&4.96&4.79\\
$\tiny{^{139}}$La&&--8.41&4.87&&4.97&4.80\\
\hline
$\tiny{^{209}_{~\Lambda}}$Pb&$1s_{1/2}$&--7.96&5.53&4.01&5.67&5.46\\
$\tiny{^{209}_{~\Lambda}}$Pb&$1p_{3/2}$&--7.94&5.53&4.72&5.67&5.47\\
$\tiny{^{208}}$Pb&&--7.87&5.53&&5.68&5.47\\
\end{longtable}
\end{indented}
\end{center}

In table 3, we list the calculated binding energy per baryon E/A
(MeV), rms charge radius $r_c$, and rms radii of the $\Lambda$, the
neutron and proton distributions ($r_\Lambda$, $r_n$, $r_p$,
respectively), for the $1s_{1/2}$ and $1p_{3/2}$ $\Lambda$
configurations. As a comparison, we also give these quantities for
normal finite nuclei.

\begin{figure}[tbp]
\begin{center}
\includegraphics[width=14cm,height=6.5cm]{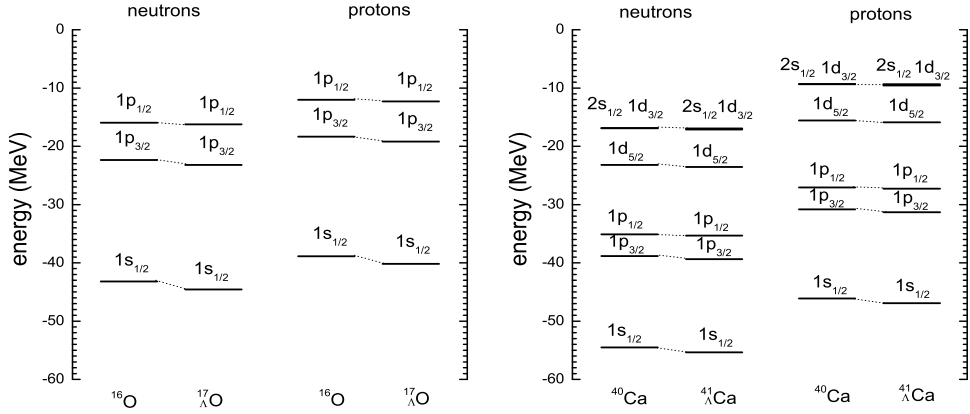}\end{center}
\caption{Nucleon single-particle energies for $\tiny{^{16}}$O,
 $\tiny{^{40}}$Ca, $\tiny{^{17}_{~\Lambda}}$O and
$\tiny{^{41}_{~\Lambda}}$Ca for the $1s_{1/2}$ $\Lambda$ state. The using parameter set is FSU.}
\label{fig3}
\end{figure}

\begin{figure}[tbp]
\begin{center}
\includegraphics[width=15cm,height=13cm]{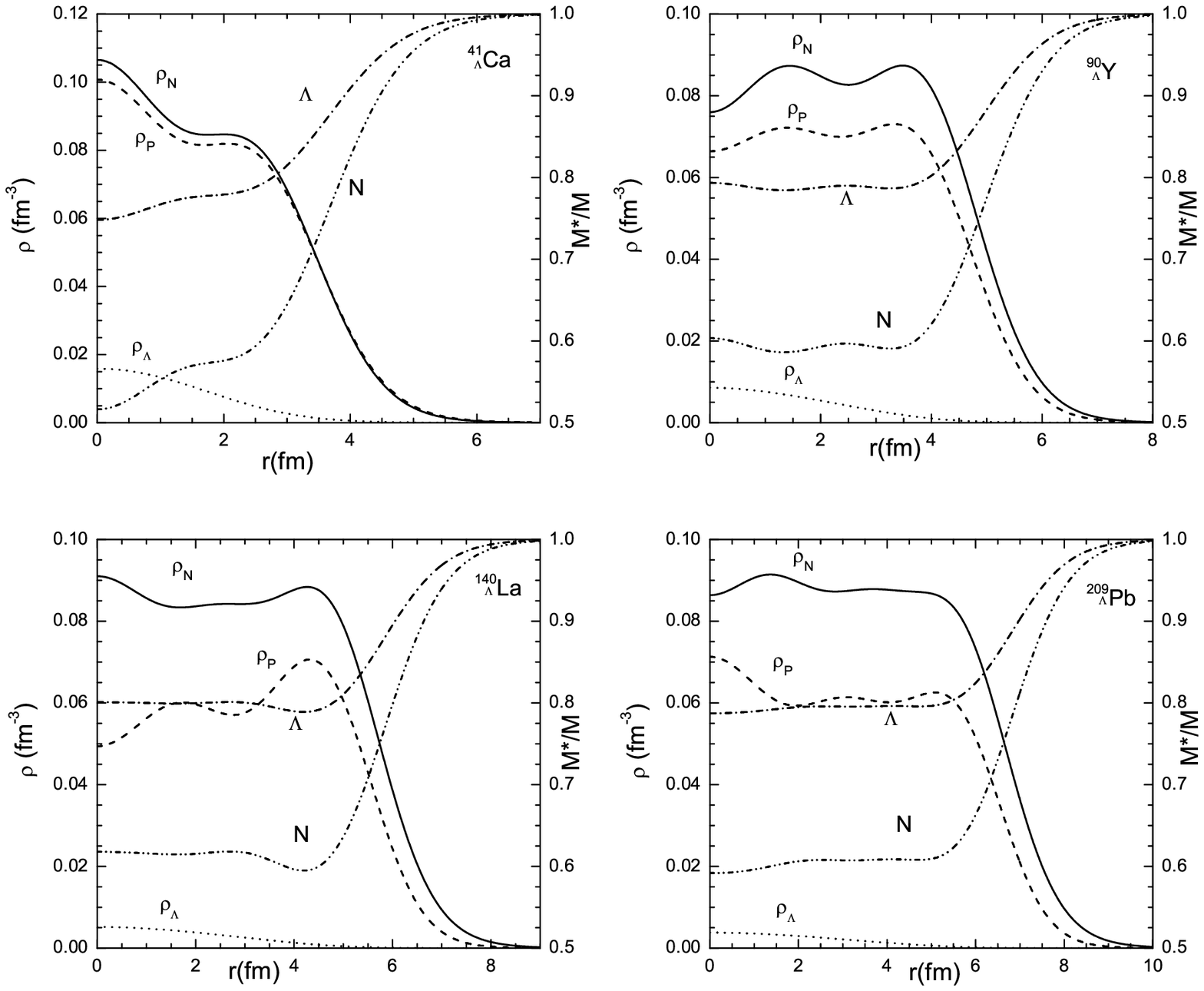}\end{center}
\caption{Calculated proton, neutron, $\Lambda$ hyperon densites, and
effective masses of the nucleon denoted by, N, and the $\Lambda$
hyperon denoted by, $\Lambda$, in hypernuclei for
$\tiny{^{41}_{~\Lambda}}$Ca, $\tiny{^{90}_{~\Lambda}}$Y,
$\tiny{^{140}_{~~\Lambda}}$La, $\tiny{^{209}_{~~\Lambda}}$Pb. All
cases are for the $1s_{1/2}$ $\Lambda$ state. The using parameter set is FSU.} \label{fig4}
\end{figure}
It is seen that the $\Lambda$ hyperon has a weak influence on the
rms charge radius, and rms radii of the nucleon. In addition, the
rms radius of the $\Lambda$ in $1s_{1/2}$ state tends to be bigger as
the nuclear number increases. Regarding the effects of the $\Lambda$
on the core nucleons, we also show in Fig.3 the nucleon single
particle energies for $\tiny{^{16}}$O, $\tiny{^{17}_{~\Lambda}}$O,
$\tiny{^{40}}$Ca and $\tiny{^{41}_{~\Lambda}}$Ca for $1s_{1/2}$
$\Lambda$ state. The existence of the $\Lambda$ makes the scalar and
baryon densities larger, and the scalar and vector potentials become
stronger. As a result, the binding energies of nucleons in
$\tiny{^{17}_{~\Lambda}}$O, $\tiny{^{41}_{~\Lambda}}$Ca are more
deeper than those of $\tiny{^{16}}$O, $\tiny{^{40}}$Ca,
respectively.

Finally, we show the effective masses of the nucleon and $\Lambda$
as well as the baryon densities calculated for
$\tiny{^{41}_{~\Lambda}}$Ca, $\tiny{^{90}_{~\Lambda}}$Y,
$\tiny{^{140}_{~~\Lambda}}$La, $\tiny{^{209}_{~~\Lambda}}$Pb for the
$1s_{1/2}$ $\Lambda$ state in Fig.4. We see that the effective
masses in these hypernuclei behave in a similar manner as the
distance r from the center of each nucleus increases
\cite{Tsushima}, and with increasing nuclear number, the $\Lambda$
hyperon density near the center of the hypernuclei becomes smaller.
This is mainly because that the rms radius of the $\Lambda$ in $1s_{1/2}$
state becomes bigger as the nuclear number increases.

\section{summary}
In the present work, the properties of some typical spherical
hypernuclei have been systemically investigated in the RMF model
with FSU parameter set as well as the NL3 and NL3$^*$ ones.
We discussed the influence of the tensor coupling between $\omega$
and $\Lambda$ on the single-$\Lambda$ hypernuclei.
In our calculation, we obtained that the $\delta_\Lambda$ of 1p orbit for
$\tiny{^{16}_{~\Lambda}}$O is --0.23 MeV with the presence of tensor coupling,
in contrast to the calculated value of --1.41 MeV
when tensor coupling is switched off.
From the calculation, we find that the tensor
potential is comparable in magnitude with scalar and vector
potentials for the $\Lambda$ hyperon in $\Lambda$ hypernuclei, but
it has a negative sign, which causes the spin-orbit term for
$\Lambda$ to be very small.
It is also shown that the tensor coupling 
leads to the inversion of energy levels between some 
states (corresponding to the pseudospin doublets) for $\Lambda$
hyperon. Our calculation also showed that the $\Lambda$ hyperon has
a weak influence on the rms charge radius, and rms radii of the
nucleon density distribution. The $\Lambda$ binding energies in
single-$\Lambda$ hypernuclei in our calculation are in good
agreement with the available experimental data. So we can see that
the parameter set FSU is as successful as the NL3 and NL3$^*$ in
reproducing the properties of single-$\Lambda$ hypernuclei.

\ack We thank Prof. W. L. Qian for valuable discussion and his
advice on the language of this manuscript. This work was supported
by the National Natural Science Foundation of China (Grants No.
11035001, No. 10735010, No. 10975072, No. 11120101005, and No.
11105072), by the 973 National Major State Basic Research and
Development of China (Grants No. 2007CB815004 and No. 2010CB327803),
by CAS Knowledge Innovation Project No. KJCX2-SW-N02, by Research
Fund of Doctoral Point, Grant No. 20100091110028, and by the Project
Funded by the Priority Academic Program Development of Jiangsu
Higher Education Institutions (PAPD).


\section*{References}
\begin{thebibliography}{10}

\bibitem{Ring}  Ring P 1996 {\it Prog. Part. Nucl. Phys.} {\bf 37} 193.

\bibitem{Lalazissis} Lalazissis G A, Sharma M M, Ring P and Gambhir Y K 1996 {\it Nucl. Phys.} A {\bf 608} 202.

\bibitem{Sugahara1}   Sugahara Y and Toki H 1994 {\it Nucl. Phys.} A {\bf 579} 557.

\bibitem{Ren}  Ren Z Z, Zhu Z Y, Cai Y H and Xu G O 1996 {\it Phys. Lett.} B {\bf 380} 241.

\bibitem{Rufa} Rufa M {\it et al.} 1990 {\it Phys. Rev.} C {\bf 42} 2469.

\bibitem{Mares1}  Glendenning N K, Von-Eiff D, Haft M, Lenske H and Weigel M K 1993 {\it Phys. Rev.} C {\bf 48} 889.

\bibitem{Win}  Win M T, Hagino K and  Koike T 2011 {\it Phys. Rev.} C {\bf 83} 014301.

\bibitem{Vretenar}  Vretenar D,  P\"{o}schl W,  Lalazissis G A  and  Ring P 1998 {\it Phys. Rev.} C {\bf 57} R1060.

\bibitem{Tsushima1}   Tsushima K, Saito K, Haidenbauer J and  Thomas A W 1998 {\it Nucl. Phys.} A {\bf 630} 691.

\bibitem{Shen1}   Shen H and Toki H 2002 {\it Nucl. Phys.} A {\bf 707} 469.

\bibitem{Star} The STAR Collaboration 2010 {\it Science} {\bf 328} 58 .

\bibitem{Lalazissis1} Lalazissis G A,  K\"{o}nig J  and  Ring P 1997 {\it Phys. Rev.} C {\bf 55} 540.

\bibitem{Lalazissis2} Lalazissis G A, Raman S and Ring P 1999 {\it At. Data Nucl. Data Tables} {\bf 71} 1.

\bibitem{Lalazissis3} Lalazissis G A {\it et al.} 2009 {\it Phys.Lett.} B
 {\bf 671} 36.

\bibitem{Todd-Rutel}  Todd-Rutel B G and  Piekarewicz J 2005 {\it Phys. Rev. Lett.} {\bf 95} 122501.

\bibitem{Horowitz} Horowitz C J and Piekarewicz J 2001  {\it Phys. Rev. Lett.} {\bf 86} 5647.

\bibitem{Sheng}  Sheng Z Q,  Ren Z Z and Jiang W Z 2010 {\it Nucl. Phys.} A {\bf 832} 49.

\bibitem{Wu}  Wu C and Ren Z Z 2011 {\it Phys. Rev.} C {\bf 83} 025805.

\bibitem{Jennings}  Jennings B K 1990 {\it Phys. Lett.} B {\bf 246} 325.

\bibitem{Toki}  Toki H, Hirata D, Sugahara Y, Sumiyoshi K and Tanihata I 1995 {\it Nucl. Phys.} A {\bf 588} c357.

\bibitem{Gambhir}  Gambhir Y K, Ring P and Thimet A 1990 {\it Ann. Phys. (N.Y.)} {\bf 198} 132.

\bibitem{Fattoyev}  Fattoyev F J, Horowitz C J, Piekarewicz J and Shen G 2010 {\it Phys. Rev.} C {\bf 82} 055803.

\bibitem{Shen} Shen H, Yang F and Toki H 2006 {\it Prog.Theor.Phys.} {\bf 115} 325.

\bibitem{Keil}  Keil C M, Hofmann F and Lenske H 2000 {\it Phys. Rev.} C {\bf 61} 064309.

\bibitem{Usmani} Usmani Q N and Bodmer A R 1999 {\it Phys. Rev.} C {\bf 60} 055215.

\bibitem{Hashimoto1}  Hashimoto O and Tamura H 2006 {\it Prog. Part. Nucl. Phys.} {\bf 57} 564.



\bibitem{Sugahara} Sugahara Y and Toki H  {\it Prog. Theo. Phys.} 1994 {\bf 92} 803.

\bibitem{Millener} Millener D J, Dover C B  and  Gal A 1988 {\it Phys. Rev.} C {\bf 38} 2700.

\bibitem{Mares}  Mare\v{s} J and  Jennings B K 1994 {\it Phys. Rev.} C {\bf 49} 2472.

\bibitem{Ma}  Ma Z Y, Speth J, Krewald S, Chen B Q and Reuber A 1996 {\it Nucl. Phys.} A {\bf 608} 305.

\bibitem{Tsushima}  Tsushima K, Saito K and Thomas A W 1997 {\it Phys. Lett.} B {\bf 411} 9.

\end{thebibliography}
\end{document}